\documentclass[aps,preprint]{revtex4}%
\usepackage{amsfonts}
\usepackage{amsmath}
\usepackage{amssymb}
\usepackage{graphicx}%
\begin{document}
\preprint{CTP-SCU/2012002}
\title{Homogeneous Field and WKB Approximation In Deformed Quantum Mechanics with
Minimal Length}
\author{Jun Tao$^{a}$}
\email{taojun@scu.edu.cn}
\author{Peng Wang$^{a}$}
\email{pengw@scu.edu.cn}
\author{Haitang Yang$^{a,b}$}
\email{hyanga@scu.edu.cn}
\affiliation{$^{a}$Center for Theoretical Physics, College of Physical Science and
Technology, Sichuan University, Chengdu, 610064, PR China}
\affiliation{$^{b}$Kavli Institute for Theoretical Physics China (KITPC), Chinese Academy
of Sciences, Beijing 100080, China}

\begin{abstract}
{}In the framework of the deformed quantum mechanics with a minimal length, we
consider the motion of a non-relativistic particle in a homogeneous external
field. We find the integral representation for the physically acceptable wave
function in the position representation.\ Using the method of steepest
descent, we obtain the asymptotic expansions of the wave function at large
positive and negative arguments. We then employ the leading asymptotic
expressions to derive the WKB connection formula, which proceeds from
classically forbidden region to classically allowed one through a turning
point. By the WKB connection formula, we prove the Bohr-Sommerfeld
quantization rule up to $\mathcal{O}\left(  \beta^{2}\right)  $. We also show
that, if the slope of the potential at a turning point is too steep, the WKB
connection formula is no longer valid around the turning point. The effects of
the minimal length on the classical motions are investigated using the
Hamilton-Jacobi method. We also use the Bohr-Sommerfeld quantization to study
statistical physics in deformed spaces with the minimal length.

\end{abstract}
\maketitle
\tableofcontents

\section{Introduction}

One of the predictions shared by various quantum theories of gravity is the
existence of a minimal observable length. For example, this fundamental
minimal length scale could arise in the framework of the\ string
theory\cite{Veneziano1986EPL199,DJGross1988NPB407, DAmati1989PLB41}. For a
review of a minimal length in quantum gravity, see \cite{Garay1995IJMPA145}.
Some realizations of the minimal length from various scenarios have been
proposed. Specifically, one of the most popular models is the generalized
uncertainty principle (GUP)\cite{Maggiore1993PLB319,Kempf1995PRD1108}, derived
from the modified fundamental commutation relation
\begin{equation}
\lbrack X,P]=i\hbar(1+\beta P^{2}), \label{1dGUP}%
\end{equation}
where $\beta=\beta_{0}\ell_{p}^{2}/\hbar^{2}=\beta_{0}/c^{2}m_{pl}^{2}$,
$m_{pl}$ is the Planck mass, $\ell_{p}$ is the Planck length, and $\beta_{0}$
is a dimensionless parameter. For a review of the GUP, see
\cite{Hossenfelder:2012jw}. With this generalization, one can easily derive
the generalized uncertainty principle (GUP)
\begin{equation}
\Delta X\Delta P\geq\frac{\hbar}{2}[1+\beta(\Delta P)^{2}]. \label{2dGUP}%
\end{equation}
This in turn gives the minimal measurable length
\begin{equation}
\Delta X\geq\Delta_{\text{min}}=\hbar\sqrt{\beta}=\sqrt{\beta_{0}}\ell_{p}.
\label{GUPshixian-2}%
\end{equation}
Eqn. (\ref{1dGUP}) is the simplest model where only the minimal uncertainty in
position is taken into account while the momentum can be infinite. When
incorporating the GUP into quantum field theory, one needs to generalize
deformed commutation relations to include time. However, the existence of the
minimal length could lead to Planck scale departures from Lorentz symmetry.
Therefore, the corresponding deformed commutation relations are not Lorentz
invariant and give rise to some version of the doubly special
relativity\cite{Hossenfelder2003,Camacho2003,Maziashvili2013,Berger2011}.

In this paper we consider one dimensional non-relativistic quantum mechanics
with the deformed commutation relation (\ref{1dGUP}). To implement the
deformed commutators (\ref{1dGUP}), one
defines\cite{Kempf1995PRD1108,Das2008PRL221301}%
\begin{equation}
X=X_{0}\text{, }P=P_{0}\left(  1+\frac{\beta}{3}P_{0}^{2}\right)  ,
\end{equation}
where $[X_{0},P_{0}]=i\hbar$, the usual canonical operators. One can easily
show that to the first order of $\beta$, eqn. (\ref{1dGUP}) is guaranteed.
Henceforth, terms of $\mathcal{O}\left(  \beta^{2}\right)  $ and higher are
neglected in the remainder of the paper. For a quantum system described by%
\begin{equation}
H=\frac{P^{2}}{2m}+V\left(  X\right)  ,
\end{equation}
the Hamiltonians can be written as
\begin{equation}
H=H_{0}+H_{1}+\mathcal{O}\left(  \beta^{2}\right)  ,
\end{equation}
where $H_{0}=\frac{P_{0}^{2}}{2m}+V\left(  X_{0}\right)  $ and $H_{1}%
=\frac{2\beta}{3}P_{0}^{2}$. Furthermore, one can adopt the momentum
representation%
\begin{equation}
X_{0}=i\hbar\frac{\partial}{\partial p}\text{, }P_{0}=p,
\end{equation}
or the position representation%
\begin{equation}
X_{0}=x\text{, }P_{0}=\frac{\hbar}{i}\frac{\partial}{\partial x}\text{.}%
\end{equation}
The momentum representation is very handy in the discussions of certain
problems, such as the harmonic oscillator\cite{Chang2002PRD125027}, the
Coulomb potential\cite{Yao2003PLB42,Fityo2006JPAM2149} and the gravitational
well\cite{Brau2006PRD036002,Pedram2011JHEP093,Pedram2012IJTP1910}. Recently, a
wide class of problems, like scattering from a barrier or a particle in a
square well\cite{Das2009PLB497,Das2011PRD044013,Pedram2012PRD024016} are
discussed in position representation. Moreover, in the position
representation, it is much easier to derive and discuss WKB approximation in
the deformed quantum mechanics analogously to in the ordinary quantum
mechanics\cite{Fityo2006JPAM387}. Thus, we adopt the position representation
in this paper. In the position representation, the deformed stationary
Schrodinger equation is%
\begin{equation}
\frac{d^{2}\psi\left(  x\right)  }{dx^{2}}-\ell_{\beta}^{2}\frac{d^{4}%
\psi\left(  x\right)  }{dx^{4}}+\frac{2m\left(  E-V\left(  x\right)  \right)
}{\hbar^{2}}\psi\left(  x\right)  =0.
\end{equation}
where we define $\ell_{\beta}^{2}=\frac{2}{3}\hbar^{2}\beta$ for later convenience.

Although, the homogeneous field potential $V\left(  X\right)  =FX$ is not
studied so intensively as the quantum well, it has an important application in
theoretical physics. In the ordinary quantum mechanics, the solutions to the
Schrodinger equation with the linear potential are Airy functions, which are
essential to derive the WKB connection formulas through a turning point. This
motivates us to study the linear potential in the deformed quantum mechanics.

In the deformed quantum mechanics with minimal length, the WKB approximation
formulas are obtained in \cite{Fityo2006JPAM387}. In addition, the deformed
Bohr-Sommerfeld quantization is used to acquire energy spectra of bound states
in various
potentials\cite{Pedram2012PRD024016,Fityo2006JPAM2149,Fityo2006JPAM387,Pedram2012PLB317,Ching2012arviv1642}%
. Therefore, it is interesting to derive the WKB connection formulas through a
turning point and rigorously verify the Bohr-Sommerfeld quantization rule
claimed before, which are presented in our paper. Besides, we find that, if
the slope of the potential is too steep at a turning point, the WKB connection
algorithm fails around the turning point. This is not unexpected because, if
one makes linear approximation to the potential around such a turning point
for asymptotic matching, the corrections to the wave functions due to the
Hamiltonian $H_{1}$ become dominant before one reaches the WKB valid region.

This paper is organized as follows: In section \ref{sec:scheqn} we give the
integral representation of the physically acceptable wave function of the
homogeneous field and its leading asymptotic behavior at large positive value
of $\rho$. In section \ref{sec:asymexp}, we obtain the asymptotic expansions
of the physically acceptable wave function at both large positive and large
negative values of $\rho$. Section \ref{sec:WKB} is devoted to deriving the
WKB connection formula and the related discussions and applications. In
section \ref{sec:conclusion}, we offer a summary and conclusion.

\section{Deformed Schrodinger Equation}

\label{sec:scheqn}

Let us consider one-dimensional motion of a particle in a homogenous field,
specifically in a field with the potential $V\left(  X\right)  =FX$. Here we
take the direction of the force along the axis of $-x$ and let $F$ be the
force exerting on the particle in the field. As discussed in the introduction,
the deformed Schrodinger equation for this scenario is%
\begin{equation}
\frac{d^{2}\psi\left(  x\right)  }{dx^{2}}-\ell_{\beta}^{2}\frac{d^{4}%
\psi\left(  x\right)  }{dx^{4}}+\frac{2m\left(  E-Fx\right)  }{\hbar^{2}}%
\psi\left(  x\right)  =0. \label{Schrodinger's Eq}%
\end{equation}
In order to solve eqn. (\ref{Schrodinger's Eq}), a new dimensionless variable
$\rho$ is introduced as%
\begin{equation}
\rho=\left(  x-\frac{E}{F}\right)  \left(  2mF/\hbar^{2}\right)  ^{\frac{1}%
{3}}.
\end{equation}
Eqn. (\ref{Schrodinger's Eq}) then becomes%
\begin{equation}
-\alpha^{2}\psi^{\left(  4\right)  }+\psi^{\prime\prime}-\rho\psi=0.
\label{differential Eq}%
\end{equation}
where we define another dimensionless variable $\alpha^{2}=\ell_{\beta}%
^{2}\left(  2mF/\hbar^{2}\right)  ^{\frac{2}{3}}$ and the derivatives are in
terms of the new variable $\rho$. The linear differential equation
(\ref{differential Eq}) is quartic and then there are four linearly
independent solutions. We will shortly show that only one of them is
physically acceptable.

\subsection{Physically Acceptable Solution}

The condition $\beta P^{2}\ll1$ validating our effective GUP model implies%
\begin{equation}
\beta\left\langle x\right\vert P^{2}\left\vert \psi\right\rangle
\ll\left\langle x | \psi\right\rangle \Longrightarrow
\alpha^{2}\left\vert \psi^{\prime\prime}\left(  \rho\right)  \right\vert
\ll\left\vert \psi\left(  \rho\right)  \right\vert . \label{asymrelation}%
\end{equation}
This condition is also expected in the momentum space. Since the GUP model is
only valid below the energy scale $\beta^{-\frac{1}{2}}$, the momentum
spectrum of the state $\left\vert \psi\right\rangle $ should be greatly
suppressed around the scale $\beta^{-\frac{1}{2}}$. It also leads to the
condition (\ref{asymrelation}). Moreover, the condition (\ref{asymrelation})
and eqn. (\ref{differential Eq}) give%
\begin{equation}
\alpha^{2}\left\vert \rho\right\vert \ll1. \label{physicalcondition}%
\end{equation}
In other words, our GUP model, which is an effective model, is valid only when
the condition (\ref{physicalcondition}) holds. Considering that the Compton
wavelength of a particle should be much larger than $\hbar\sqrt{\beta}$ or
$\ell_{\beta}$ in the GUP model, one can also obtain the condition
(\ref{physicalcondition}) in the classical allowed region where $\rho<0$. In a
field with the potential $V\left(  x\right)  $, the kinematics energy of a
non-relativistic particle is $E-V\left(  x\right)  $ and its momentum is
$\sqrt{2m\left(  E-V\left(  x\right)  \right)  }$. Therefore, the fact that
the Compton wavelength of the particle $\lambda_{c}=\frac{\hbar}%
{\sqrt{2m\left(  E-V\left(  x\right)  \right)  }}$ is much larger than
$\ell_{\beta}$ yields $\alpha^{2}\left\vert \rho\right\vert \ll1$. In the
remainder of our paper except subsubsection \ref{stp}, we assume $\alpha\ll1$
which is useful to derive WKB connection formula around a smooth tuning point.
One needs to consider $\alpha\gtrapprox1$ scenario only when it comes to the
WKB connection around a sharp turning point.\ 

We notice that $E<V$ for $\rho>0$. The wave function $\psi$ is then
exponentially damped for large positive value of $\rho$.\ Thus, one needs to
evaluate asymptotic values of $\psi\left(  \rho\right)  $ at large positive
value of $\rho$ to find physically acceptable solution to eqn.
(\ref{differential Eq}). Note that, only when $\alpha\ll1$, one can analyze
asymptotic behavior of $\psi\left(  \rho\right)  $ at large positive value of
$\rho$ in the physically acceptable region where $\alpha^{2}\left\vert
\rho\right\vert \ll1$. To determine the leading behavior of $\psi\left(
\rho\right)  $ at large positive value of $\rho$, we make the exponential
substitution $\psi\left(  \rho\right)  =e^{s\left(  \rho\right)  }$ and then
obtain for eqn. (\ref{differential Eq})
\begin{equation}
s^{^{\prime\prime}}+s^{\prime2}-\rho-\alpha^{2}\left[  s^{\left(  4\right)
}+6s^{\prime2}s^{\prime\prime}+3s^{\prime\prime2}+4s^{\prime}s^{\left(
3\right)  }+s^{\prime4}\right]  =0. \label{sdifferentialEq}%
\end{equation}
Eqn. (\ref{sdifferentialEq}) is as difficult to solve as eqn.
(\ref{differential Eq}). Here our strategy to find the asymptotic behavior of
$\psi\left(  \rho\right)  $ from eqn. (\ref{sdifferentialEq}) is as
follows\cite{CoolBook}:

\begin{description}
\item[(a)] We neglect all terms appearing small and approximate the exact
differential equation with the asymptotic one.

\item[(b)] We solve the resulting equation and check that the solution is
consistent with approximations made in step (a).
\end{description}

It is usually true that higher derivative terms than $s^{\prime}$ are
discarded in step (a). Therefore, we reduce eqn. (\ref{sdifferentialEq}) to
the asymptotic differential equation%
\begin{equation}
s^{\prime2}-\alpha^{2}s^{\prime4}\sim\rho. \label{asymdiffeq}%
\end{equation}
Solving eqn. $\left(  \ref{asymdiffeq}\right)  $ gives four solutions for
$s^{\prime}$, two of which are discarded considering $\beta P^{2}\ll1$. Taking
asymptotic relation (\ref{asymrelation}) into account, one can further reduce
the quartic equation (\ref{asymdiffeq}) to a quadratic equation%
\begin{equation}
s^{\prime2}\sim\rho.
\end{equation}
which has only two solutions for $s^{\prime}$. The two solutions are
$s^{\prime}\sim\pm\sqrt{\rho}$, and, therefore,%
\begin{equation}
\psi\left(  \rho\right)  \sim\exp\left(  \pm\frac{2}{3}\rho^{\frac{3}{2}%
}\right)  ,\text{ at large positive value of }\rho,
\end{equation}
where $-$ is for the physically acceptable solution. It is easy to check that
the solution $s^{\prime}\sim\pm\sqrt{\rho}$ satisfy the assumptions%
\[
s^{^{\prime\prime}}\text{, }s^{\prime2}s^{\prime\prime}\text{, }s^{\left(
3\right)  }\text{, }s^{\prime\prime2}\text{, }s^{\prime}s^{\left(  3\right)
}\text{ and }s^{\left(  4\right)  }\ll\rho,\text{ }%
\]
as long as $\rho\gg1$.

It is interesting to note that the two discarded solutions of eqn.
(\ref{asymdiffeq}) are
\begin{equation}
s^{\prime}\sim\pm\frac{\sqrt{1+\sqrt{1-4\alpha^{2}\rho}}}{\sqrt{2\alpha}},
\end{equation}
which become $s^{\prime}\sim\pm\frac{1}{\sqrt{\alpha}}$ when $\alpha^{2}%
\rho\ll1.$ The resulting wave functions are $\psi\left(  \rho\right)  \sim
\exp\left(  \pm\frac{\rho}{\sqrt{\alpha}}\right)  $. They are not physical
states since they fail to satisfy the condition (\ref{asymrelation}). One can
also see that these two solutions are discarded according to the low-momentum
consistency condition in \cite{Ching2012arviv1519}. In summary, assuming
$\alpha\ll1$, we find that the leading asymptotic behavior of the physically
acceptable solution is $\exp\left(  -\frac{2}{3}\rho^{\frac{3}{2}}\right)  $
for $\rho\gg1$. In addition, we only analyze the solution in the region
$\alpha^{2}\left\vert \rho\right\vert \ll1$ where the GUP model is valid.

\subsection{Integral Representation}

The differential equation eqn. (\ref{differential Eq}) can be solved by
Laplace's method. Please refer to mathematical appendices of \cite{Landau} for
more details. Define the polynomials%
\begin{equation}
P\left(  t\right)  =-\alpha^{2}t^{4}+t^{2},\text{ }Q\left(  t\right)  =-1,
\end{equation}
and the function%
\begin{equation}
Z\left(  t\right)  =\frac{1}{Q\left(  t\right)  }\exp\left(  \int
\frac{P\left(  t\right)  }{Q\left(  t\right)  }dt\right)  =-\exp\left(
\frac{\alpha^{2}t^{5}}{5}-\frac{t^{3}}{3}\right)  .
\end{equation}
Integral representations of the solutions to eqn. (\ref{differential Eq}) are
then given by%
\begin{align}
\psi\left(  \rho\right)   &  =-\int_{C}\exp\left(  \rho t\right)  Z\left(
t\right)  dt\nonumber\\
&  =\int_{C}\exp\left(  \rho t+\frac{\alpha^{2}t^{5}}{5}-\frac{t^{3}}%
{3}\right)  dt, \label{solution}%
\end{align}
where the contour $C$ is chosen so that the integral is finite and non-zero
and the function%
\begin{equation}
V\left(  t\right)  =\exp\left(  \rho t+\frac{\alpha^{2}t^{5}}{5}-\frac{t^{3}%
}{3}\right)  ,
\end{equation}
vanishes at endpoints of $C$ since the integrand of eqn. (\ref{solution}) is
entire on the complex plane of $t$. Now that $\exp\left(  xt+\frac{\alpha
^{2}t^{5}}{5}-\frac{t^{3}}{3}\right)  \sim\exp\left(  \frac{\alpha^{2}t^{5}%
}{5}\right)  $for large $t$, we need to begin and end the contour $C$ in
sectors for which $\cos5\theta<0$ (setting $t=|t|e^{i\theta}$). There are five
such sectors, specifically
\begin{equation}
\theta\in\Theta_{n}\equiv\left[  \frac{2n\pi+\frac{\pi}{2}}{5},\frac
{2n\pi+\frac{3\pi}{2}}{5}\right]  ,\text{ }n=0,1,2,3,4,5.
\end{equation}
Therefore, any contour which originates at one of them and terminates at
another yields a solution to eqn. (\ref{differential Eq}). One could then find
four linearly independent functions of the form%
\begin{equation}
I_{i}\left(  \rho\right)  =\int_{C_{i}}\exp\left(  \rho t+\frac{\alpha
^{2}t^{5}}{5}-\frac{t^{3}}{3}\right)  dt. \label{foursolutions}%
\end{equation}
The asymptotic expression for $I_{i}\left(  \rho\right)  $ for large values of
$\rho$ is obtained by evaluating the integral eqn. (\ref{foursolutions}) by
the method of steepest descents.

\section{Asymptotic Expansion}

\label{sec:asymexp}

First we briefly review the method of steepest descent to introduce some
useful formulas. This technique is very powerful to calculate integrals of the
form%
\begin{equation}
I\left(  \rho\right)  =\int_{C}g\left(  z\right)  e^{\rho f\left(  z\right)
}dz,
\end{equation}
where $C$ is a contour in the complex plane and $g\left(  z\right)  $ and
$f\left(  z\right)  $ are analytic functions. The parameter $\rho$ is real and
we are usually interested in the behaviors of $I\left(  \rho\right)  $ as
$\rho\rightarrow\pm\infty$. The key step of the method of steepest descent is
applying Cauchy's theorem to deform the contours $C$ to the contours
consisting of steepest descent paths and other paths joining endpoints of two
different steepest descent paths if necessary. Usually, the joining paths are
chosen to make negligible contributions to $I\left(  \rho\right)  $. It is
easy to show that $\operatorname{Im}f\left(  z\right)  $ is constant along
steepest descent paths. When a steepest descent contour passes through a
saddle point $z_{0}$ where $f^{\prime}\left(  z_{0}\right)  =0$, $f\left(
z\right)  $ and $g\left(  z\right)  $ are expanded around $z_{0}$ and Watson's
lemma is used to determine asymptotic behaviors of $I\left(  \rho\right)  $.
Specifically, consider a contour $C$ through a saddle point $z_{0}$. A new
variable $\tau$ is introduced as $\tau=f\left(  z\right)  -f\left(
z_{0}\right)  $ to calculate $I\left(  \rho\right)  $. The saddle point
$z_{0}$ divides the contour $C$ into two contours $C_{1}$ and $C_{2}$.
Generally, $\tau$ monotonically increases from $-\infty$ to zero along one
contour, say $C_{1}$ and monotonically decreases from zero to $-\infty$ along
$C_{2}$. Thus, the integral becomes%
\begin{equation}
I\left(  \rho\right)  =\exp\left[  \rho f\left(  z_{0}\right)  \right]
\left[  \int_{-\infty}^{0}g\left(  \tau\right)  \exp\left[  \rho\tau\right]
\frac{dz}{d\tau}|_{C_{1}}d\tau+\int_{0}^{-\infty}g\left(  \tau\right)
\exp\left[  \rho\tau\right]  \frac{dz}{d\tau}|_{C_{2}}d\tau\right]  .
\end{equation}

The physically acceptable solution can be represented by an integral%
\begin{equation}
I\left(  \rho\right)  =\int_{C}\exp\left(  \rho t+\frac{\alpha^{2}t^{5}}%
{5}-\frac{t^{3}}{3}\right)  dt,
\end{equation}
where $C$ is any contour which ranges from $t=\exp\left(  -\frac{3\pi i}%
{5}\right)  \infty$ to $t=\exp\left(  \frac{3\pi i}{5}\right)  \infty$. In
fact, as we show later in the section for positive $\rho$, there exists a
steepest descent contour from $t=\exp\left(  -\frac{3\pi i}{5}\right)  \infty$
to $t=\exp\left(  \frac{3\pi i}{5}\right)  \infty$, which $C$ can be deformed
to. Moreover, the integral on such a steepest descent contour yields the
required asymptotic behavior of $I\left(  \rho\right)  $ at large positive
value of $\rho$. Here the exponent in the integrand has movable saddle points.
Making the change of variables $t=\left\vert \rho\right\vert ^{\frac{1}{2}}s$,
one gets
\begin{align}
I\left(  \rho\right)   &  =\left\vert \rho\right\vert ^{\frac{1}{2}}\int
_{\exp\left(  -\frac{3\pi i}{5}\right)  \infty}^{\exp\left(  \frac{3\pi i}%
{5}\right)  \infty}\exp\left[  \left\vert \rho\right\vert ^{\frac{3}{2}%
}\left(  \pm s+\frac{as^{5}}{5}-\frac{s^{3}}{3}\right)  \right]  ds\nonumber\\
&  \equiv\left\vert \rho\right\vert ^{\frac{1}{2}}\int_{\exp\left(
-\frac{3\pi i}{5}\right)  \infty}^{\exp\left(  \frac{3\pi i}{5}\right)
\infty}\exp\left[  \left\vert \rho\right\vert ^{\frac{3}{2}}f_{\pm}\left(
s\right)  \right]  ds,
\end{align}
where $+$ for $\rho>0$ and $-$ for $\rho<0$ and $a=\alpha^{2}\left\vert
\rho\right\vert \ll1$ in the physical region.

\subsection{Large Positive $\rho$}

For $\rho>0$, we have%
\begin{equation}
f_{+}\left(  s\right)  =s+\frac{as^{5}}{5}-\frac{s^{3}}{3}.
\end{equation}
There are four saddle points given by $f_{+}^{\prime}\left(  s\right)  =0$ at%
\begin{equation}
s=\pm\lambda_{+}\equiv\pm\frac{\sqrt{1-\sqrt{1-4a}}}{\sqrt{2a}}\text{ and
}s=\pm\eta_{+}\equiv\pm\frac{\sqrt{1+\sqrt{1-4a}}}{\sqrt{2a}}.
\end{equation}
Our goal now is to find a steepest descent contour emerging from
$s=\exp\left(  -\frac{3\pi i}{5}\right)  \infty$ to $s=\exp\left(  \frac{3\pi
i}{5}\right)  \infty$. We will show that such a contour passes through
$s=-\lambda_{+}$. To find the contour we substitute $s=u+iv$ and identify the
real and imaginary parts of $f_{+}\left(  s\right)  $%
\begin{gather}
f_{+}\left(  s\right)  =u\left(  1-\frac{u^{2}}{3}+\frac{au^{4}}{5}%
+v^{2}-2au^{2}v^{2}+av^{4}\right) \nonumber\\
+iv\left(  1-u^{2}+au^{4}+\frac{v^{2}}{3}-2au^{2}v^{2}+\frac{av^{4}}%
{5}\right)  .
\end{gather}
Since $\operatorname{Im}f_{+}\left(  -\lambda_{+}\right)  =0$, the
constant-pahse contours passing through $s=-\lambda_{+}$ must satisfy%
\begin{equation}
v\left(  1-u^{2}+au^{4}+\frac{v^{2}}{3}-2au^{2}v^{2}+\frac{av^{4}}{5}\right)
=0.
\end{equation}
Therefore, one of the constant-phase contours passing through $s=-\lambda_{+}$
is%
\[
C:-\frac{1}{\sqrt{2a}}\sqrt{1+2av^{2}-\sqrt{1-4a+\frac{8}{3}av^{2}+\frac
{16}{5}a^{2}v^{4}}}+iv,\text{ for}-\infty<v<\infty\text{,}%
\]
which is a steepest descent contour. In fact, around the saddle point
$s=-\lambda_{+}$, one finds on the contour $C$%
\begin{equation}
s\sim-\lambda_{+}+bv^{2}+iv,
\end{equation}
and hence,%
\begin{equation}
f_{+}\left(  s\right)  =f_{+}\left(  -\lambda_{+}\right)  -\frac{v^{2}}%
{2}f_{+}^{\prime\prime}\left(  -\lambda_{+}\right)  +\mathcal{O}\left(
v^{3}\right)  ,
\end{equation}
where $b$ is a positive real number. Since $f_{+}^{\prime\prime}\left(
-\lambda_{+}\right)  $ is real and positive, the contour $C$ is indeed a
steepest descent contour. Note that $C$ goes to $s=\exp\left(  -\frac{3\pi
i}{5}\right)  \infty$ as $v\rightarrow-\infty$ and $s=\exp\left(  \frac{3\pi
i}{5}\right)  \infty$ as $v\rightarrow\infty$. In order to evaluate asymptotic
expansion of $I\left(  \rho\right)  $, we break up the contour $C$ into
$C_{1}$ and $C_{2}$, corresponding to above and below of $s=-\lambda_{+}$.
Define%
\begin{equation}
\tau=f_{+}\left(  s\right)  -f_{+}\left(  -\lambda_{+}\right)  , \label{stau1}%
\end{equation}
where $\tau$ monotonically decreases from zero to $-\infty$ as one moves away
from $s=-\lambda_{+}$ along $C_{1}$ to $s=\exp\left(  \frac{3\pi i}{5}\right)
\infty$ and along $C_{2}$ to $s=\exp\left(  -\frac{3\pi i}{5}\right)  \infty$,
respectively. Since $f_{+}^{\prime}\left(  -\lambda_{+}\right)  =0,$ the
expression for $s$ in terms of $\tau$ can be expressed as a power series of
$\sqrt{-\tau}$. Then, noting that $-\tau=\left(  \pm\sqrt{-\tau}\right)  ^{2}%
$, one has%
\begin{equation}
s=-\lambda_{+}+\sum_{j=1}^{\infty}a_{j}\left(  \pm\sqrt{-\tau}\right)  ^{j},
\label{s-tau1}%
\end{equation}
where $a_{i}$ can be obtained by substituting eqn. (\ref{s-tau1}) into eqn.
(\ref{stau1}) and equating powers of $\sqrt{-\tau}$ on both sides of the
equations. It is easy to find%
\begin{equation}
a_{1}=i\sqrt{\frac{2}{\left\vert f_{+}^{\prime\prime}\left(  -\lambda
_{+}\right)  \right\vert }},
\end{equation}
where one finds $\operatorname{Im}a_{1}>0$. The contour $C_{1}$ is in the
second quadrant and hence, $+$ sign is chosen in eqn. (\ref{s-tau1}) for
$C_{1}$. Therefore,%
\begin{gather}
\rho^{\frac{1}{2}}\int_{C_{1}}\exp\left[  \rho^{\frac{3}{2}}f_{+}\left(
s\right)  \right]  ds=\rho^{\frac{1}{2}}\exp\left[  \rho^{\frac{3}{2}}%
f_{+}\left(  -\lambda_{+}\right)  \right]  \int_{0}^{-\infty}\exp\left(
\rho^{\frac{3}{2}}\tau\right)  \frac{ds}{d\tau}d\tau\nonumber\\
\sim\exp\left[  \rho^{\frac{3}{2}}f_{+}\left(  -\lambda_{+}\right)  \right]
\sum_{j=1}^{\infty}\frac{ja_{j}}{2\rho^{\frac{3j-2}{4}}}\Gamma\left(  \frac
{j}{2}\right)  . \label{asymplusc1}%
\end{gather}
For the contour segment $C_{2}$, the sign of $\sqrt{\tau}$ occurring in eqn.
(\ref{s-tau1}) has to be reversed. Moreover, the limit of integration on
$C_{2}$ in the variable $\tau$ ranges from $-\infty$ to $0$. Thus,%
\begin{equation}
\rho^{\frac{1}{2}}\int_{C_{1}}\exp\left[  \rho^{\frac{3}{2}}f_{+}\left(
s\right)  \right]  ds\sim-\exp\left[  \rho^{\frac{3}{2}}f_{+}\left(
-\lambda_{+}\right)  \right]  \sum_{j=1}^{\infty}\frac{\left(  -1\right)
^{j}ja_{j}}{2\rho^{\frac{3j-2}{4}}}\Gamma\left(  \frac{j}{2}\right)  .
\label{asymplusc2}%
\end{equation}
Combining eqn. (\ref{asymplusc1}) and eqn. (\ref{asymplusc2}), we easily find%
\begin{equation}
I\left(  1\ll\rho\ll\alpha^{-2}\right)  \sim\frac{\exp\left[  \rho^{\frac
{3}{2}}f_{+}\left(  -\lambda_{+}\right)  \right]  }{\rho^{\frac{1}{4}}}%
\sum_{j=0}^{\infty}\frac{\left(  2j+1\right)  a_{2j+1}}{\rho^{\frac{3j}{2}}%
}\Gamma\left(  j+\frac{1}{2}\right)  .
\end{equation}

\subsection{Large Negative $\rho$}

As for $\rho<0$, the exponent in the integrand of $I\left(  \rho\right)  $ is%
\begin{equation}
f_{-}\left(  s\right)  =-s+\frac{as^{5}}{5}-\frac{s^{3}}{3}.
\end{equation}
Thus, one as well finds four saddle points given by $f_{+}^{\prime}\left(
s\right)  =0$%
\begin{equation}
s=\pm\lambda_{-}\equiv\pm\frac{\sqrt{1-\sqrt{1+4a}}}{\sqrt{2a}}\text{ and
}s=\pm\eta_{-}\equiv\pm\frac{\sqrt{1+\sqrt{1+4a}}}{\sqrt{2a}}.
\label{saddlepointminus}%
\end{equation}
As before, our objective is to find steepest descent contours passing through
the saddle point(s) in eqn. (\ref{saddlepointminus}) that emerges from
$s=\exp\left(  -\frac{3\pi i}{5}\right)  \infty$ to $s=\exp\left(  \frac{3\pi
i}{5}\right)  \infty$. Substituting $s=u+iv$, we obtain the real and imaginary
parts of $f_{-}\left(  s\right)  $%
\begin{gather}
f_{-}\left(  s\right)  =u\left(  -1-\frac{u^{2}}{3}+\frac{au^{4}}{5}%
+v^{2}-2au^{2}v^{2}+av^{4}\right) \nonumber\\
+iv\left(  -1-u^{2}+au^{4}+\frac{v^{2}}{3}-2au^{2}v^{2}+\frac{av^{4}}%
{5}\right)  .
\end{gather}
We have already shown that only one steepest descent contour passing through
$s=-\lambda_{+}$ is sufficient to evaluate asymptotic behavior of $I\left(
\rho\right)  $ for large and positive $\rho$. However for large and negative
$\rho$, things are a little bit more complicated. Instead of one steepest
descent contour, it turns out that we need three steepest descent contours
passing through $\pm\lambda_{-}$ and $\eta_{-}$, respectively, to connect two
endpoints at $s=\exp\left(  \pm\frac{3\pi i}{5}\right)  \infty$.

First consider the steepest descent contour through $s=-\lambda_{-}$. Since
$f_{+}\left(  -\lambda_{-}\right)  $ is a pure imaginary number, the steepest
descent contour must satisfy%
\begin{equation}
iv\left(  -1-u^{2}+au^{4}+\frac{v^{2}}{3}-2au^{2}v^{2}+\frac{av^{4}}%
{5}\right)  =f_{-}\left(  -\lambda_{-}\right)  .
\end{equation}
Solutions to the last equation give us a constant phase contour $C_{-\lambda
_{-}}$ passing through $s=-\lambda_{-}$, which emanates from $s=\exp\left(
-\frac{3\pi i}{5}\right)  \infty$ and finally approaches $s=\exp\left(
-\frac{\pi i}{5}\right)  \infty$. The contour $C_{-\lambda_{-}}$ actually is
composed of three segments as%
\begin{align*}
C_{-\lambda_{-},1}  &  :-\frac{1}{\sqrt{2a}}\sqrt{1+2av^{2}-\sqrt
{F_{-\lambda_{-}}\left(  v\right)  }}+iv,\text{ for }-\infty
<v<-\operatorname{Im}\lambda_{-},\\
C_{-\lambda_{-},2}  &  :\frac{1}{\sqrt{2a}}\sqrt{1+2av^{2}-\sqrt
{F_{-\lambda_{-}}\left(  v\right)  }}+iv,\text{ for }-\operatorname{Im}%
\lambda_{-}<v<-v_{0},\\
C_{-\lambda_{-},3}  &  :\frac{1}{\sqrt{2a}}\sqrt{1+2av^{2}+\sqrt
{F_{-\lambda_{-}}\left(  v\right)  }}+iv,\text{ for }-v_{0}>v>-\infty,
\end{align*}
where we define%
\[
F_{\pm\lambda_{-}}\left(  v\right)  =1+4a+\frac{8}{3}av^{2}+\frac{16}{5}%
a^{2}v^{4}+\frac{4af_{-}\left(  \pm\lambda_{-}\right)  }{iv},
\]
and $v_{0}$ is a solution to $F_{+\lambda_{-}}\left(  v\right)  =0$ that
satisfies $0<v_{0}\ll1$. It is straightforward to verify that, along
$C_{-\lambda_{-}}$, $\operatorname{Re}f_{-}\left(  s\right)  $ monotonically
increases from $-\infty$ to $0$ as one moves from $s=\exp\left(  -\frac{3\pi
i}{5}\right)  \infty$ to $s=-\lambda_{-}$ and then monotonically decreases
from $0$ to $-\infty$ as one moves away from $s=-\lambda_{-}$ to
$s=\exp\left(  -\frac{\pi i}{5}\right)  \infty$. Hence, the contour
$C_{-\lambda_{-}}$ is indeed a the steepest descent contour passing through
$s=-\lambda_{-}$. Now we calculate the contour integral on $C_{-\lambda_{-}}$.
Introduce%
\begin{equation}
\tau=f_{-}\left(  s\right)  -f_{-}\left(  -\lambda_{-}\right)  ,
\label{s-tau2}%
\end{equation}
which $\tau$ is real on $C_{-\lambda_{-}}$ and varies from $-\infty$ to zero
and then to $-\infty$ along $C_{-\lambda_{-}}$. Then, one has
\begin{equation}
s=-\lambda_{-}+\sum_{j=1}^{\infty}b_{j}\left(  \pm\sqrt{-\tau}\right)  ^{j},
\label{stau2}%
\end{equation}
where $b_{i}$ can be obtained by substituting eqn. (\ref{s-tau2}) into eqn.
(\ref{stau2}). One easily gets%
\begin{equation}
b_{1}=\exp\left(  \frac{\pi}{4}i\right)  \sqrt{\frac{2}{\left\vert
f_{-}^{\prime\prime}\left(  -\lambda_{+}\right)  \right\vert }}.
\end{equation}
Since $\operatorname{Re}\exp\left(  \frac{\pi}{4}i\right)  >0$, one has
$-\sqrt{-\tau}$ for $C_{-\lambda_{-},1}$\ and $\sqrt{-\tau}$ for
$C_{-\lambda_{-},2}+C_{-\lambda_{-},3}$ in eqn. (\ref{stau2}). Therefore,%
\begin{gather}
\left\vert \rho\right\vert ^{\frac{1}{2}}\int_{C_{-\lambda_{-}}}\exp\left[
\left\vert \rho\right\vert ^{\frac{3}{2}}f_{-}\left(  s\right)  \right]
ds\nonumber\\
\sim2\rho^{\frac{1}{2}}\exp\left[  \left\vert \rho\right\vert ^{\frac{3}{2}%
}f_{-}\left(  -\lambda_{-}\right)  \right]  \int_{0}^{-\infty}\exp\left(
\left\vert \rho\right\vert ^{\frac{3}{2}}\tau\right)  \sum_{j=0}^{\infty
}\left(  2j+1\right)  b_{j}\left(  \sqrt{-\tau}\right)  ^{2j}d\sqrt{-\tau
}\label{Iminus1}\\
=\frac{\exp\left[  \left\vert \rho\right\vert ^{\frac{3}{2}}f_{-}\left(
-\lambda_{-}\right)  \right]  }{\left\vert \rho\right\vert ^{\frac{1}{4}}}%
\sum_{j=0}^{\infty}\frac{\left(  2j+1\right)  b_{j}}{\left\vert \rho
\right\vert ^{\frac{3j}{2}}}\Gamma\left(  j+\frac{1}{2}\right)  .\nonumber
\end{gather}

Analogously, one can readily write down a constant phase contour
$C_{+\lambda_{-}}$ passing through $s=-\lambda_{-}$, which starts from
$s=\exp\left(  \frac{\pi i}{5}\right)  \infty$ and ends at $s=\exp\left(
\frac{3\pi i}{5}\right)  \infty$. As before, $C_{+\lambda_{-}}$ consists of
three segments%
\begin{align*}
C_{+\lambda_{-},1}  &  :\frac{1}{\sqrt{2a}}\sqrt{1+2av^{2}+\sqrt
{F_{+\lambda_{-}}\left(  v\right)  }}+iv,\text{ for }+\infty>v>v_{0},\\
C_{+\lambda_{-},2}  &  :\frac{1}{\sqrt{2a}}\sqrt{1+2av^{2}-\sqrt
{F_{+\lambda_{-}}\left(  v\right)  }}+iv,\text{ \ for }v_{0}%
>v>\operatorname{Im}\lambda_{-},\\
C_{+\lambda_{-},3}  &  :-\frac{1}{\sqrt{2a}}\sqrt{1+2av^{2}-\sqrt
{F_{+\lambda_{-}}\left(  v\right)  }}+iv,\text{ for }\operatorname{Im}%
\lambda_{-}<v<+\infty.
\end{align*}
\ \ \ It is also straightforward to verify that $C_{+\lambda_{-}}$ is a
steepest descent contour as well. Setting%
\begin{equation}
\tau=f_{-}\left(  s\right)  -f_{-}\left(  \lambda_{-}\right)  , \label{s-tau3}%
\end{equation}
one finds $\tau$ is real on $C_{+\lambda_{-}}$ and varies from $-\infty$ to
zero and then to $-\infty$ along $C_{+\lambda_{-}}$. Note that $f_{+}\left(
s\right)  $ is an odd function and $\lambda_{-}^{\ast}=-\lambda_{-}$. Taking
complex conjugate of both sides of eqn. (\ref{s-tau3}), one then has on
$C_{+\lambda_{-}}$
\begin{equation}
s=\lambda_{-}+\sum_{j=1}^{\infty}b_{j}^{\ast}\left(  \pm\sqrt{-\tau}\right)
^{j}. \label{stau3}%
\end{equation}
Since $\operatorname{Re}b_{1}^{\ast}>0$, one has $\sqrt{-\tau}$ for
$C_{+\lambda_{-},1}+C_{+\lambda_{-},2}$\ and $-\sqrt{-\tau}$ for
$C_{+\lambda_{-},3}$ in eqn. (\ref{stau3}). Therefore,%
\begin{equation}
\left\vert \rho\right\vert ^{\frac{1}{2}}\int_{C_{+\lambda_{-}}}\exp\left[
\left\vert \rho\right\vert ^{\frac{3}{2}}f_{-}\left(  s\right)  \right]
ds\sim-\frac{\exp\left[  \left\vert \rho\right\vert ^{\frac{3}{2}}f_{-}\left(
\lambda_{-}\right)  \right]  }{\left\vert \rho\right\vert ^{\frac{1}{4}}}%
\sum_{j=0}^{\infty}\frac{b_{j}^{\ast}\left(  2j+1\right)  }{\left\vert
\rho\right\vert ^{\frac{3j}{2}}}\Gamma\left(  j+\frac{1}{2}\right)  .
\label{Iminus2}%
\end{equation}

Since the values of $\operatorname{Im}f_{-}\left(  s\right)  $ are different
on $C_{\pm\lambda_{-}}$, it is obvious that we need a third contour which
joins $C_{\pm\lambda_{-}}$ up at $s=\exp\left(  \pm\frac{\pi i}{5}\right)
\infty$, respectively. Here, we consider a constant phase contour $C_{\eta
_{-}}$ connecting $s=\exp\left(  -\frac{\pi i}{5}\right)  \infty$ to
$\exp\left(  \frac{\pi i}{5}\right)  \infty$ that passes through $\eta_{-}$.
Since $\operatorname{Im}f\left(  s\right)  =\operatorname{Im}f_{-}\left(
\eta_{-}\right)  =0$ on the contour $C_{\eta_{-}},$ one finds%
\[
C_{\eta_{-}}:\frac{1}{\sqrt{2a}}\sqrt{1+2av^{2}+\sqrt{1+4a+\frac{8}{3}%
av^{2}+\frac{16}{5}a^{2}v^{4}}}+iv,\text{ for }-\infty<v<+\infty,
\]
is a curve of steepest descent. On $C_{\eta_{-}}$, define%
\begin{equation}
\tau=f_{-}\left(  s\right)  -f_{-}\left(  \eta_{-}\right)  ,
\end{equation}
which is real on $C_{\eta_{-}}$ and varies from $-\infty$ to zero and then to
$-\infty$ along $C_{\eta_{-}}$. Then, one finds%
\begin{equation}
s=\lambda_{-}+\sum_{j=1}^{\infty}c_{j}\left(  \pm\sqrt{-\tau}\right)  ^{j},
\label{stau4}%
\end{equation}
and
\begin{equation}
c_{1}=i\sqrt{\frac{2}{\left\vert f_{-}^{\prime\prime}\left(  \eta_{-}\right)
\right\vert }}.
\end{equation}
Similarly, we break up the contour $C_{\eta_{-}}$ into $C_{\eta_{-},1}$ and
$C_{\eta_{-},2}$, corresponding to above and below of $s=\eta_{-}$ with
$\sqrt{-\tau}$ for $C_{\eta_{-},1}$ and $-\sqrt{-\tau}$ for $C_{\eta_{-},2}$
in eqn. (\ref{stau4}). Thus,%
\begin{equation}
\left\vert \rho\right\vert ^{\frac{1}{2}}\int_{C_{\eta_{-}}}\exp\left[
\left\vert \rho\right\vert ^{\frac{3}{2}}f_{-}\left(  s\right)  \right]
ds\sim\frac{\exp\left[  \left\vert \rho\right\vert ^{\frac{3}{2}}f_{-}\left(
\eta_{-}\right)  \right]  }{\left\vert \rho\right\vert ^{\frac{1}{4}}}%
\sum_{j=0}^{\infty}\frac{\left(  2j+1\right)  c_{2j+1}}{\left\vert
\rho\right\vert ^{\frac{3j}{2}}}\Gamma\left(  j+\frac{1}{2}\right)  .
\label{Iminus3}%
\end{equation}

Note that although paths $C_{\eta_{-}}$ and $C_{\pm\lambda_{-}}$ never join up
at $s=\exp\left(  \pm\frac{\pi i}{5}\right)  \infty$, the integrand
$\exp\left[  f_{-}\left(  s\right)  \right]  \sim\exp\left(  \frac{a\left\vert
\rho\right\vert ^{\frac{3}{2}}}{5}s^{5}\right)  $ tends to zero exponentially.
Therefore, there is no contribution from a connecting path from $C_{\eta_{-}}$
and $C_{\pm\lambda_{-}}$ at a distance $R$ from the origin in the limit
$R\rightarrow\infty$. As a result, the integral $I\left(  \rho\right)  $
equals to the sum of three contour integrals on the different steepest descent
curves $C_{\eta_{-}}$ and $C_{\pm\lambda_{-}}$. Combining eqn. (\ref{Iminus1}%
), eqn. (\ref{Iminus2}) and eqn. (\ref{Iminus3}) gives the full asymptotic
expansion of $I\left(  \rho\right)  $ for large and negative $\rho$%
\begin{align}
I\left(  -1\gg\rho\gg-\alpha^{-2}\right)   &  \sim2i\sum_{j=0}^{\infty}%
\frac{\operatorname{Im}\left(  \exp\left[  -\left\vert \rho\right\vert
^{\frac{3}{2}}f_{-}\left(  \lambda_{-}\right)  \right]  b_{j}\right)
}{\left\vert \rho\right\vert ^{\frac{1}{4}}}\frac{\left(  2j+1\right)
}{\left\vert \rho\right\vert ^{\frac{3j}{2}}}\Gamma\left(  j+\frac{1}%
{2}\right) \nonumber\\
&  +\frac{\exp\left[  \left\vert \rho\right\vert ^{\frac{3}{2}}f_{-}\left(
\eta_{-}\right)  \right]  }{\left\vert \rho\right\vert ^{\frac{1}{4}}}%
\sum_{j=0}^{\infty}\frac{\left(  2j+1\right)  c_{2j+1}}{\left\vert
\rho\right\vert ^{\frac{3j}{2}}}\Gamma\left(  j+\frac{1}{2}\right)  .
\end{align}

\section{WKB Approximation}

\label{sec:WKB}

The authors of \cite{Fityo2006JPAM387} find the WKB approximation in deformed
space with minimal length. In \cite{Fityo2006JPAM387}, they consider the
deformed commutation relation%
\begin{equation}
\left[  X,P\right]  =i\hbar f\left(  P\right)  , \label{eq:generalGUP}%
\end{equation}
where $f\left(  P\right)  $ is an arbitrary function of $P$. In our paper, we
set $f\left(  P\right)  =1+\beta P^{2}$. Defining $P\left(  p\right)  $%
\begin{equation}
\frac{dP\left(  p\right)  }{dp}=f\left(  P\right)  ,
\end{equation}
and $p\left(  P\right)  $ an inverse function of $P\left(  p\right)  ,$ they
find the physical-optics approximation to the solution of the deformed
Schrodinger equation%
\begin{equation}
P^{2}\left(  \frac{\hbar}{i}\frac{d}{dx}\right)  \psi\left(  x\right)
+2m\left[  V\left(  x\right)  -E\right]  \psi\left(  x\right)  =0,
\end{equation}
is%
\begin{equation}
\psi\left(  x\right)  =\frac{1}{\sqrt{\left\vert P\left(  x\right)  f\left(
P\left(  x\right)  \right)  \right\vert }}\left(  C_{1}\exp\left[  \frac
{i}{\hbar}\int^{x}p\left(  x\right)  dx\right]  +C_{2}\exp\left[  -\frac
{i}{\hbar}\int^{x}p\left(  x\right)  dx\right]  \right)  , \label{WKB}%
\end{equation}
where $P\left(  x\right)  =\sqrt{2m\left(  E-V\left(  x\right)  \right)  }$
and $p\left(  x\right)  =p\left(  P\left(  x\right)  \right)  $ in eqn.
(\ref{WKB}). It is also shown there that, if eqn. (\ref{WKB}) is valid, the
condition
\begin{equation}
\left\vert P^{2}\left(  x\right)  \right\vert \gg\hbar\left\vert \frac{d}%
{dx}P\left(  x\right)  f\left(  P\left(  x\right)  \right)  \right\vert ,
\label{Condition}%
\end{equation}
has to be satisfied. However, the condition eqn. (\ref{Condition}) fails near
a turning point where $P\left(  x\right)  =0$. Thus, if we want to determine
bound state energies, we need to be able to match wave functions at the
turning points. Here we considers a potential $V\left(  x\right)  $ with its
classical turning point located at $x=0$. A linear approximation to the
potential $V\left(  x\right)  $\ near the turning point $x=0$ is
\begin{equation}
V\left(  x\right)  \approx V\left(  0\right)  +Fx, \label{linearV}%
\end{equation}
where $F=V^{\prime}\left(  0\right)  $. The linearized potential
(\ref{linearV}) is discussed in the previous two sections. Our discussion
shows that the parameter $\alpha=\ell_{\beta}\left(  2m\left\vert F\right\vert
/\hbar^{2}\right)  ^{\frac{1}{3}}$ plays an important role in analyzing
asymptotic behaviors of the solutions. When $\alpha\ll1$, the physically
acceptable solution can exist at large argument $\rho$ while the condition
(\ref{physicalcondition}) still holds. Accordingly, a turning points is called
a smooth one if $\alpha=\ell_{\beta}\left(  2m\left\vert F\right\vert
/\hbar^{2}\right)  ^{\frac{1}{3}}\ll1$. Otherwise, it is called a sharp
turning point.

\subsection{WKB Connection through a Smooth Turning Point}

Now we want to match WKB wave functions at a smooth turning point in the
deformed space with $f\left(  P\right)  =1+\beta P^{2}$ up to $\mathcal{O}%
\left(  \beta\right)  $. Suppose $x=0$ is a smooth turning point, which means
$V\left(  0\right)  =E$, and $V>E$ for all $x>0$. The region to the left of
the turning point is classically forbidden where the wave function must be
damped and become zero at infinity. Thus, far from $x=0$, the wave function
has the form%
\begin{equation}
\psi\left(  x\right)  =\frac{1}{\sqrt{\left\vert P\left(  x\right)  f\left(
P\left(  x\right)  \right)  \right\vert }}C\exp\left[  -\frac{1}{\hbar
}\left\vert \int_{0}^{x}p\left(  x\right)  dx\right\vert \right]
,\text{\ \ for }x>0. \label{damped}%
\end{equation}
To the right of the turning point, the wave function is given by%
\begin{equation}
\psi\left(  x\right)  =\frac{1}{\sqrt{\left\vert P\left(  x\right)  f\left(
P\left(  x\right)  \right)  \right\vert }}\left(  C_{1}\exp\left[  \frac
{i}{\hbar}\int_{0}^{x}p\left(  x\right)  dx\right]  +C_{2}\exp\left[
-\frac{i}{\hbar}\int_{0}^{x}p\left(  x\right)  dx\right]  \right)
,\text{\ for\ }x<0. \label{oscillate}%
\end{equation}
Around the turning point, $x$ is small and $P\left(  x\right)  \sim\sqrt
{2mF}\sqrt{-x}$. In this region, we may approximate eqn. (\ref{damped}) and
eqn. (\ref{oscillate}) by%
\begin{equation}
\psi\left(  x\right)  \approx\left(  2mF\hbar\right)  ^{-\frac{1}{3}}%
x^{-\frac{1}{4}}\left(  1+\frac{3a}{4}+\mathcal{O}\left(  a^{2}\right)
\right)  C\exp\left[  -\frac{2}{3\hbar}x^{\frac{3}{2}}\left(  1+\frac{3}%
{10}a+\mathcal{O}\left(  a^{2}\right)  \right)  \right]  ,\text{ for }x>0,
\label{approxpos}%
\end{equation}%
\begin{gather}
\psi\left(  x\right)  =\left(  2mF\hbar\right)  ^{-\frac{1}{3}}x^{-\frac{1}%
{4}}\left(  1-\frac{3a}{4}+\mathcal{O}\left(  a^{2}\right)  \right)
\nonumber\\
\left(  C_{1}\exp\left[  \frac{2i}{3\hbar}\left\vert x\right\vert ^{\frac
{3}{2}}\left(  1-\frac{3}{10}a+\mathcal{O}\left(  a^{2}\right)  \right)
\right]  +C_{2}\exp\left[  -\frac{2i}{3\hbar}\left\vert x\right\vert
^{\frac{3}{2}}\left(  1-\frac{3}{10}a+\mathcal{O}\left(  a^{2}\right)
\right)  \right]  \right)  \text{, for }x<0. \label{approxneg}%
\end{gather}
The criteria (\ref{Condition}) for validity of the WKB approximation is
satisfied if
\begin{equation}
\left\vert x\right\vert \gg\left(  \frac{2mF}{\hbar^{2}}\right)  ^{-\frac
{1}{3}}, \label{WKBcondition}%
\end{equation}
where we neglect $\beta P^{2}$ in derivation. On the other hand, when the
potential is linearized around the turning point $x=0$, the Schrodinger
equation becomes%
\begin{equation}
\frac{d^{2}\psi\left(  x\right)  }{dx^{2}}-\ell_{\beta}^{2}\frac{d^{4}%
\psi\left(  x\right)  }{dx^{4}}-\frac{2m\mu x}{\hbar^{2}}\psi\left(  x\right)
\approx0, \label{linearschrodinger}%
\end{equation}
where $\beta=\frac{3\ell_{\beta}^{2}}{2\hbar^{2}}$. To solve the approximate
differential equation, we make the substitution%
\begin{equation}
\rho=x\left(  2mF/\hbar^{2}\right)  ^{\frac{1}{3}}.
\end{equation}
In terms of $\rho,$ the solution to eqn. (\ref{linearschrodinger}) which
matches eqn. (\ref{approxpos}$)$ and eqn. (\ref{approxneg}) in two different
limits is actually $I\left(  \rho\right)  $ calculated in the section
\ref{sec:asymexp}. Specifically, the solution is%
\begin{equation}
\psi\left(  x\right)  =DI\left(  \rho\right)  =DI\left(  x\left(
2mF/\hbar^{2}\right)  ^{\frac{1}{3}}\right)  , \label{linearSol}%
\end{equation}
where $D$ is a constant to be determined by asymptotic matching. It is easily
shown from (\ref{WKBcondition}) that there exists overlap regions where both
WKB approximation and eqn. (\ref{linearschrodinger}) hold. In the overlap
regions, one finds $\left\vert \rho\right\vert \gg1$ and $\left\vert
x\right\vert \ll1$. Therefore, we approximate $I\left(  \rho\right)  $ by its
leading asymptotic behaviors for large argument in the the overlap regions.
The appropriate formulas are
\begin{align}
I\left(  1\ll\rho\ll\alpha^{-2}\right)   &  \sim\frac{i\sqrt{\pi}\left(
1+\frac{3a}{4}+\mathcal{O}\left(  a^{2}\right)  \right)  }{\rho^{\frac{1}{4}}%
}\exp\left[  -\frac{2\rho^{\frac{3}{2}}}{3}\left(  1+\frac{3a}{10}%
+\mathcal{O}\left(  a^{2}\right)  \right)  \right]  ,\label{Ipositive}\\
I\left(  -1\gg\rho\gg-\alpha^{-2}\right)   &  \sim\frac{2i\sqrt{\pi}\left(
1-\frac{3a}{4}+\mathcal{O}\left(  a^{2}\right)  \right)  }{\left\vert
\rho\right\vert ^{\frac{1}{4}}}\sin\left[  \frac{2\left\vert \rho\right\vert
^{\frac{3}{2}}}{3}\left(  1-\frac{3a}{10}+\mathcal{O}\left(  a^{2}\right)
\right)  +\frac{\pi}{4}\right]  , \label{Inegative}%
\end{align}
where $\alpha\ll1$ for a smooth turning point and $a=\ell_{\beta}^{2}\left(
2mF/\hbar^{2}\right)  ^{\frac{2}{3}}\left\vert \rho\right\vert \ll1$ as
required by the condition (\ref{asymrelation}). Requiring that eqn.
(\ref{Ipositive}) and eqn. (\ref{Inegative}) match eqn. (\ref{approxpos}) and
eqn. (\ref{approxneg}) in the overlap region, respectively, gives
$C_{1}=-iCe^{i\pi/4}$ and $C_{2}=iCe^{i\pi/4}$ up to $\mathcal{O}\left(
\beta\right)  $. In summary, in the overlap region, we find WKB solutions and
the asymptotic values of the solution to the Schrodinger equation with a
linear approximation to the potential $V\left(  x\right)  $. Then, by making
eqn. (\ref{Ipositive}) and eqn. (\ref{Inegative}) match eqn. (\ref{approxpos})
and eqn. (\ref{approxneg}) respectively, the WKB connection formula with the
deformed commutator $\left[  X,P\right]  =i\hbar\left(  1+\beta P^{2}\right)
$ is obtained up to $\mathcal{O}\left(  \beta\right)  $. The connection
formula around a smooth turning point is put in a way that%
\begin{align}
&  \frac{C}{\sqrt{\left\vert P\left(  x\right)  f\left(  P\left(  x\right)
\right)  \right\vert }}\exp\left(  -\frac{1}{\hbar}\left\vert \int_{0}%
^{x}p\left(  x\right)  dx\right\vert \right) \nonumber\\
&  \rightarrow\frac{2C}{\sqrt{\left\vert P\left(  x\right)  f\left(  P\left(
x\right)  \right)  \right\vert }}\sin\left(  \frac{1}{\hbar}\int_{0}%
^{x}p\left(  x\right)  dx+\frac{\pi}{4}\right)  \text{, up to }\mathcal{O}%
\left(  \beta^{2}\right)  , \label{connection}%
\end{align}
which is directional, just as in ordinary quantum mechanics\cite{CoolBook}.
The analysis always proceeds from classically forbidden region to classically
allowed one. For bound states, the uniqueness of the wave function in the
classically allowed region leads to the Bohr-Sommerfeld quantization condition%
\begin{equation}
\int_{a}^{b}p\left(  x\right)  dx=\left(  n+\frac{1}{2}\right)  \pi
\hbar\text{, up to }\mathcal{O}\left(  \beta^{2}\right)  , \label{bsquant}%
\end{equation}
where $a$ and $b$ are two smooth turning points for the potential $V\left(
x\right)  $. Notice that although eqn. (\ref{bsquant}) is claimed in
\cite{Fityo2006JPAM387}, one still needs to obtain the connection formula to
derive eqn. (\ref{bsquant}) rigorously, which is not presented in
\cite{Fityo2006JPAM387}.

\subsection{Discussion}

\subsubsection{Sharp Turning Point}

\label{stp}

Near a sharp turning point $x=0$, not only the WKB approximation is no longer
valid but also matching the two WKB solutions across the turning point stops
making sense. In fact, from the previous subsection, one finds that the
asymptotic matching is valid as long as there exists an overlap region where
$1\ll\left\vert \rho\right\vert \ll\alpha^{-2}$. However, such region doesn't
exist unless $\alpha\ll1$, which means that the asymptotic matching fails
through a sharp turning point.

It can be shown, through (\ref{Condition}), WKB approximations are valid as
long as $\left\vert x\right\vert \gg\left(  2m\left\vert F\right\vert
/\hbar^{2}\right)  ^{-\frac{1}{3}}$ in the region where the potential is
approximated by a linear one. Put another way, if there exists a region where
both WKB and linear approximations are valid, one finds $\left\vert
x\right\vert \gg\left(  2m\left\vert F\right\vert /\hbar^{2}\right)
^{-\frac{1}{3}}$ for such a region. When $\left\vert x\right\vert \gg\left(
2m\left\vert F\right\vert /\hbar^{2}\right)  ^{-\frac{1}{3}}$, we have%
\begin{equation}
\left\vert \beta P^{2}\right\vert \approx\frac{\ell_{\beta}^{2}}{\left(
2m\left\vert F\right\vert /\hbar^{2}\right)  ^{\frac{2}{3}}}\frac{x}{\left(
2m\left\vert F\right\vert /\hbar^{2}\right)  ^{\frac{1}{3}}}\gg1
\end{equation}
for a sharp turning point. However, $\left\vert \beta P^{2}\right\vert \ll1$
is required by the GUP model. This means that, as moving away from the sharp
turning point, one is far beyond the region where the linear approximation to
the potential is good before reaching the WKB valid region. One might resort
to a higher order approximation to the potential and asymptotic matching in
the overlap region to find WKB connection formula through a sharp turning point.

\subsubsection{$\mathcal{O}\left(  \beta\right)  $ vs. $\mathcal{O}\left(
\hbar\right)  $}

When $\hbar$ can be regarded as a small quantity, the approximate solution to
the deformed Schrodinger equation%

\begin{equation}
\frac{d^{2}\psi\left(  x\right)  }{dx^{2}}-\frac{2\hbar^{2}\beta}{3}%
\frac{d^{4}\psi\left(  x\right)  }{dx^{4}}+\frac{2m\left(  E-V\left(
x\right)  \right)  }{\hbar^{2}}\psi\left(  x\right)  =0,
\label{wkbschrodinger}%
\end{equation}
is easy to find using WKB analysis. To be specific, the approximate solution
is expressed in an exponential power series of the form%
\begin{equation}
\psi\left(  x\right)  \sim\exp\left[  \frac{1}{\hbar}\sum_{n=0}^{\infty}%
\hbar^{n}S_{n}\left(  x\right)  \right]  .
\end{equation}
The authors of \cite{Fityo2006JPAM387} finds%
\begin{equation}
S_{1}=-\frac{1}{2}\ln\left\vert 2Pf\left(  P\right)  \right\vert .
\end{equation}
Since here $f\left(  P\right)  =1+\beta P^{2}$, we have for $S_{1}$
\begin{equation}
S_{1}\approx\ln\frac{1}{\sqrt{\left\vert P\right\vert }}-\frac{\beta}{2}%
P^{2}+\mathcal{O}\left(  \beta^{2}\right)  .
\end{equation}
Moreover, the leading order (in terms of $\beta$) of the $S_{2}$ is just the
WKB $\mathcal{O}\left(  \hbar^{2}\right)  $ correction calculated in the
ordinary quantum mechanics. Therefore, we obtain\cite{Landau}%
\begin{equation}
S_{2}\approx\frac{P^{\prime}}{4P^{2}}+\int\frac{P^{\prime2}}{8P^{3}}dx+\text{
}\mathcal{O}\left(  \beta\right)  \text{.}%
\end{equation}
If one uses WKB approximations to evaluate quantum gravity induced
corrections, say to energy levels or tunnelling rates, one may want to have%
\begin{equation}
\beta P^{2}\gtrsim\hbar S_{2}. \label{betavsh}%
\end{equation}
Otherwise, the quantum gravity correction $\left(  \sim\mathcal{O}\left(
\beta\right)  \right)  $ on the first order WKB approximation $\left(
\sim\mathcal{O}\left(  \hbar^{0}\right)  \right)  $ could be overwhelmed by
the second order WKB approximation $\left(  \sim\mathcal{O}\left(
\hbar\right)  \right)  $. Suppose $a$ is the characteristic length of the
potential $V\left(  x\right)  $, for example the width of a square-well
potential. Then we can get a rough estimate on $S_{2}$%
\begin{equation}
\hbar S_{2}\sim\frac{\hbar}{aP}\sim\frac{\lambda}{a},
\end{equation}
where $\lambda$ is the de Broglie wavelength of a particle with momentum $P$.
As a result, the condition (\ref{betavsh}) becomes%
\begin{equation}
\frac{\ell_{\beta}^{2}}{\lambda^{2}}\gtrsim\frac{\lambda}{a}\Rightarrow
\lambda\lesssim\ell_{\beta}\left(  \frac{a}{\ell_{\beta}}\right)  ^{\frac
{1}{3}}. \label{betavshconstraint}%
\end{equation}
It is interesting to note that the condition (\ref{betavshconstraint}) is a
rough estimate and a more accurate estimate could be obtained once the form of
the potential is given.

Taking into account the constraints (\ref{betavshconstraint}) on the de
Broglie wavelength $\lambda$ of a particle, one may conclude that the WKB
approximation is not a powerful tool to calculate quantum gravity corrections
unless the energy of the particle considered\ is high enough.\ However, there
is an exception if the corresponding Schrodinger equation in the ordinary
quantum mechanics can be solved exactly. In this case, $\mathcal{O}\left(
\beta\right)  $ corrections calculated on the WKB first order approximation
are just quantum gravity corrections to exact results up to $\mathcal{O}%
\left(  \beta\right)  \mathcal{O}\left(  \hbar^{0}\right)  $ even without
having (\ref{betavshconstraint}) required. For example, if we employ WKB
analysis to calculate the energy spectrum of a bound state in the deformed
space, the energy levels can be represented by a series in powers of $\hbar$%
\begin{equation}
E_{n}=\sum_{j=0}^{\infty}\hbar^{j}E_{n,j}\left(  \beta\right)  ,
\label{energylevels}%
\end{equation}
where $E_{n,j}\left(  \beta\right)  $ can be expanded in terms of $\beta$%
\begin{equation}
E_{n,j}\left(  \beta\right)  =\sum_{k=0}^{\infty}\beta^{k}E_{n,j}^{k}.
\end{equation}
If on the first order WKB approximation, one calculates $E_{n,0}\left(
\beta\right)  $ up to $\mathcal{O}\left(  \beta\right)  $%
\begin{equation}
E_{n,0}\left(  \beta\right)  =E_{n,0}^{0}+\beta E_{n,0}^{1}+\mathcal{O}\left(
\beta^{2}\right)  ,
\end{equation}
the energy levels are%
\begin{equation}
E_{n}=E_{n,0}^{0}+\beta E_{n,0}^{1}+\mathcal{O}\left(  \beta^{2}\right)
+\mathcal{O}\left(  \hbar\right)  . \label{expand}%
\end{equation}
In order to have eqn. (\ref{expand}) make sense, one requires $\beta
E_{n,0}^{1}\gtrsim\mathcal{O}\left(  \hbar\right)  $. On the other hand, if we
know the exact result $E_{n}$ with $\beta=0$, namely $E_{n}^{\left(  0\right)
}$%
\begin{equation}
E_{n}^{\left(  0\right)  }=\sum_{j=0}^{\infty}\hbar^{j}E_{n,j}^{0},
\end{equation}
eqn. (\ref{energylevels}) becomes%
\begin{equation}
E_{n}=E_{n}^{\left(  0\right)  }+\beta E_{n,0}^{1}+\mathcal{O}\left(
\hbar\right)  \mathcal{O}\left(  \beta\right)  +\mathcal{O}\left(  \beta
^{2}\right)  . \label{expand2}%
\end{equation}
Since $\mathcal{O}\left(  \hbar\right)  \mathcal{O}\left(  \beta\right)  $ is
automatically smaller than $\beta E_{0}^{1}$, eqn. (\ref{expand2}) always
makes sense as long as $\mathcal{O}\left(  \hbar\right)  \ll1$.

To illustrate our points, we use the WKB approximation to derive the energy
levels of a particle confined to the one-dimensional potential $V\left(
x\right)  =F\left\vert x\right\vert $ whose turning points are%
\begin{equation}
a=-\frac{E}{F},\text{ }b=\frac{E}{F}.
\end{equation}
The energy quantization condition (\ref{bsquant}) from first order WKB
approximation then becomes%
\begin{equation}
\ell_{F}^{-\frac{3}{2}}\int_{-\frac{E}{F}}^{\frac{E}{F}}\sqrt{\frac{E}%
{F}-\left\vert x\right\vert }dx-\frac{\ell_{\beta}^{2}}{2\ell_{F}^{\frac{9}%
{2}}}\int_{-\frac{E}{F}}^{\frac{E}{F}}\left(  \frac{E}{F}-\left\vert
x\right\vert \right)  ^{\frac{3}{2}}dx=\left(  n+\frac{1}{2}\right)
\pi+\mathcal{O}\left(  \beta^{2}\right)  ,
\end{equation}
where $\ell_{F}=\left(  \hbar^{2}/2mF\right)  ^{\frac{1}{3}}$ is the
characteristic length of the potential $V\left(  x\right)  =F\left\vert
x\right\vert $. From the last equation, we obtain%
\begin{equation}
\frac{E_{n}}{F}\approx\ell_{n}\left(  1+\frac{\ell_{\beta}^{2}\ell_{n}}%
{5\ell_{F}^{3}}+\mathcal{O}\left(  \beta^{2}\right)  +\mathcal{O}\left(
\hbar\right)  \right)  , \label{beta}%
\end{equation}
where $\ell_{n}=\ell_{F}\left[  \frac{3}{4}\left(  n+\frac{1}{2}\right)
\pi\right]  ^{\frac{2}{3}}$. What is $\mathcal{O}\left(  \hbar\right)  $? The
second order generalization of eqn. (\ref{bsquant}) with $\beta=0$ is given in
\cite{CoolBook}%
\begin{equation}
\ell_{F}^{-\frac{3}{2}}\int_{-\frac{E^{\left(  0\right)  }}{F}}^{\frac
{E^{\left(  0\right)  }}{F}}\sqrt{\frac{E^{\left(  0\right)  }}{F}-\left\vert
x\right\vert }dx+\frac{F_{F}^{\frac{3}{2}}\ell^{\frac{3}{2}}}{48E^{\left(
0\right)  \frac{3}{2}}}=\left(  n+\frac{1}{2}\right)  \pi+\mathcal{O}\left(
\hbar^{2}\right)  ,
\end{equation}
which gives%
\begin{equation}
\frac{E_{n}^{\left(  0\right)  }}{F}\approx\ell_{n}\left(  1-\frac{\ell
_{F}^{3}}{96\ell_{n}^{3}}+\mathcal{O}\left(  \hbar^{2}\right)  \right)  .
\label{h2}%
\end{equation}
We can then estimate $\mathcal{O}\left(  \hbar\right)  $ through (\ref{h2})%
\begin{equation}
\mathcal{O}\left(  \hbar\right)  \sim\frac{\ell_{F}^{3}}{\ell_{n}^{3}},
\end{equation}
which can also be easily obtained by dimensional analysis. If one wants the
first order approximation (\ref{beta}) to make sense, the second term in
(\ref{beta}) should be comparable to or larger than $\mathcal{O}\left(
\hbar\right)  $ and then one gets%
\begin{equation}
\ell_{n}\gtrsim\ell_{F}\sqrt{\frac{\ell_{F}}{\ell_{\beta}}}. \label{ln}%
\end{equation}
The de Broglie wavelength of a particle with energy $E_{n}\sim F\ell_{n}$ is
\begin{equation}
\lambda_{n}\sim\frac{\hbar}{\sqrt{2mF\ell_{n}}}\sim\frac{\ell_{F}^{\frac{3}%
{2}}}{\sqrt{\ell_{n}}}.
\end{equation}
Thus, the inequality (\ref{ln}) reads%
\begin{equation}
\lambda_{n}\lesssim\ell_{F}\left(  \frac{\ell_{\beta}}{\ell_{F}}\right)
^{\frac{1}{4}}, \label{mildconstraint}%
\end{equation}
which is much milder than (\ref{betavshconstraint}). In a practical way,
$\hbar$ and $\beta$ can be expressed in terms of $\ell_{\beta}$, $l_{F}$ and
$l_{n}$. In fact, it is easily shown that%
\begin{equation}
\mathcal{O}\left(  \hbar^{m}\right)  \sim\mathcal{O}\left(  \frac{\ell
_{F}^{3m}}{\ell_{n}^{3m}}\right)  \sim\mathcal{O}\left(  \frac{1}{n^{2m}%
}\right)  \text{\thinspace},\text{ \ }\mathcal{O}\left(  \beta^{m}\right)
\sim\mathcal{O}\left(  \frac{\ell_{\beta}^{2m}}{\ell_{F}^{2m}}\right)  .
\end{equation}

\subsection{Application}

The dimensionless number $\beta_{0}=c^{2}m_{pl}^{2}\beta=\hbar^{2}\beta
/\ell_{p}^{2}$ plays an important role when implications and applications of
non-zero minimal length are discussed. Normally, if the minimal length is
assumed to be order of the Planck length $\ell_{p}$, one has $\beta_{0}\sim1$.
In \cite{Das2008PRL221301}, based on the precision measurement of Lamb shift,
an upper bound of $\beta_{0}$ was given by $\beta_{0}<10^{36}$. The authors in
\cite{Benczik:2002tt} placed constraints on $\beta_{0}$ from the precession of
the perihelion of the Mercury, which was $\beta_{0}<10^{-66}$. However, as
pointed out in \cite{Quesne:2009vc}, the effective deformation parameter was
substantially reduced by a factor $N^{-2}$ for a macroscopic body which
consists of $N$ quarks. Thus, an upper bound on $\beta_{0}$ for quarks was
$\beta_{0}^{q}<10^{36}$. In the following, we first use the Hamilton-Jacobi
method to study the effects of the minimal length on the classical motions.
The Bohr-Sommerfeld quantization is then used to investigate statistical
physics in deformed spaces with the minimal length.

\subsubsection{Hamilton-Jacobi Method in Deformed Spaces}

In \cite{Benczik:2002tt,Quesne:2009vc}, the classical limit of deformed spaces
with the minimal length has been studied by replacing the quantum mechanical
commutator by the Poisson bracket via%
\begin{equation}
\frac{1}{i\hbar}\left[  \hat{A},\hat{B}\right]  \Rightarrow\left\{
A,B\right\}  .
\end{equation}
Alternatively, we here use Hamilton-Jacobi method to probe the classical
motion of a particle with the mass $m$ under the potential $V\left(  x\right)
$ in 1D deformed spaces.

For the deformed commutation relation $\left(  \ref{eq:generalGUP}\right)  $,
the deformed time dependent Schrodinger equation is%
\begin{equation}
\frac{P^{2}\left(  \frac{\hbar}{i}\frac{\partial}{\partial x}\right)  }%
{2m}\psi\left(  x,t\right)  +V\left(  x\right)  \psi\left(  x,t\right)
=i\hbar\frac{\partial\psi\left(  x,t\right)  }{\partial t}. \label{eq:TDSE}%
\end{equation}
Substituting the ansatz $\psi\left(  x,t\right)  =\exp\left[  \frac{iS\left(
x,t\right)  }{\hbar}\right]  $ into eqn. $\left(  \ref{eq:TDSE}\right)  $ and
taking the limit $\hbar\rightarrow0$, one finds that the leading order of eqn.
$\left(  \ref{eq:TDSE}\right)  $ gives the classical Hamilton-Jacobi equation
in deformed spaces%
\begin{equation}
\frac{P^{2}\left(  \frac{\partial S}{\partial x}\right)  }{2m}+V\left(
x\right)  +\frac{\partial S}{\partial t}=0, \label{eq:DHJE}%
\end{equation}
where $S\left(  x,t\right)  $ is the classical action. Since the potential
$V\left(  x\right)  $ does not depend explicitly on time, we can separate the
variables as%
\begin{equation}
S=W\left(  x\right)  -Et,
\end{equation}
where $E$ can be identified with the total energy. Thus, the Hamilton-Jacobi
equation becomes%
\begin{equation}
P\left(  \frac{dW}{dx}\right)  =\pm\sqrt{2m\left[  E-V\left(  x\right)
\right]  }\equiv\pm P\left(  x\right)  . \label{eq:DHJEW}%
\end{equation}
Defining $p\left(  P\right)  $ as an inverse function of $P\left(  p\right)
$, we obtain%
\begin{equation}
\frac{dW}{dx}=\pm p\left(  P\left(  x\right)  \right)  \equiv p\left(
x\right)  . \label{eq:DHJEp}%
\end{equation}
Eqn. $\left(  \ref{eq:DHJEp}\right)  $ can be integrated to%
\begin{equation}
W=\pm\int p\left(  x\right)  dx,
\end{equation}
so that%
\begin{equation}
S=\pm\int p\left(  x\right)  dx-Et.
\end{equation}
The solution for $x\left(  t\right)  $ comes from the transformation equation%
\begin{equation}
C=\frac{\partial S}{\partial E}=\pm\int\frac{\partial p\left(  x\right)
}{\partial E}dx-t=\pm\int\frac{mdx}{f\left(  P\left(  x\right)  \right)
P\left(  x\right)  }-t, \label{eq:Ceqn}%
\end{equation}
where the constant $C$ can be determined by the initial conditions and we use
\begin{equation}
\frac{\partial P\left(  x\right)  }{\partial E}=\frac{m}{P\left(  x\right)
}\text{ and }\frac{dp}{dP}=\frac{1}{f\left(  P\right)  }\text{.}%
\end{equation}
Now we focus on the case with $f\left(  P\right)  =1+\beta P^{2}$.

First we consider the motion of a particle under the homogeneous field
potential $V\left(  x\right)  =Fx$. Eqn. $\left(  \ref{eq:Ceqn}\right)  $
leads to%
\begin{equation}
\left(  1-\frac{F}{E}x\right)  \left[  1-\frac{4\beta mE}{3}\left(  1-\frac
{F}{E}x\right)  \right]  =\frac{F^{2}m}{2E}\left(  \frac{t+C}{m}\right)  ^{2},
\label{eq:yeqn}%
\end{equation}
where we neglect the terms higher than $\mathcal{O}\left(  \beta\right)  $.
Solving eqn. $\left(  \ref{eq:yeqn}\right)  $ for $x$ to $\mathcal{O}\left(
\beta^{2}\right)  $ gives
\begin{equation}
x\left(  t\right)  =\frac{E}{F}-\frac{F}{2m}\left(  t+C\right)  ^{2}%
-mE\beta\frac{F^{3}}{3Em^{2}}\left(  t+C\right)  ^{4},
\end{equation}
which indicates the initial conditions at $t=-C$ are $x\left(  -C\right)
=\frac{E}{F}\equiv x_{0}$ and $x^{\prime}\left(  -C\right)  =0.$
Differentiating both sides of eqn. $\left(  \ref{eq:yeqn}\right)  $ twice with
respect to $t$ gives the acceleration%
\begin{equation}
a\equiv\frac{d^{2}x}{dt^{2}}=-\frac{F}{m}\left[  1+\frac{8}{3}\beta
m^{2}x^{\prime2}-\frac{8}{3}\beta mF\left(  x-x_{0}\right)  \right]  .
\label{eq:aeqn}%
\end{equation}
Using $x-x_{0}=\frac{mx^{\prime2}}{2F}+\mathcal{O}\left(  \beta\right)  $, we
find from eqn. $\left(  \ref{eq:aeqn}\right)  $ that%
\begin{equation}
a=-\frac{F}{m}\left(  1+\frac{4}{3}\beta m^{2}x^{\prime2}\right)
+\mathcal{O}\left(  \beta^{2}\right)  . \label{eq:aaeqn}%
\end{equation}

The equivalence principle is crucial in the foundations of general relativity.
The weak equivalence principle is often referred to as the universality of
free fall. A measure for the breakdown of the universality of free fall is the
\textquotedblleft Eotvos ratio\textquotedblright\cite{Will:2014kxa}%
\begin{equation}
\eta\left(  A,B\right)  =\frac{2\left\vert a_{A}-a_{B}\right\vert }{\left\vert
a_{A}+a_{B}\right\vert },
\end{equation}
which quantifies the normalized difference in the gravitational accelerations
between two different bodies $A$ and $B$. In the modern torsion-balance
experiment\cite{Schlamminger:2007ht}, the \textquotedblleft Eotvos
ratio\textquotedblright\ has been found to be%
\begin{equation}
\eta\left(  \text{Be},\text{Ti}\right)  =\left(  0.3\pm1.8\right)
\times10^{-13},
\end{equation}
for the gravitational acceleration of Beryllium and Titanium towards the
Earth. In the experiment of \cite{Schlamminger:2007ht}, they used%
\begin{equation}
m_{\text{Be}}\approx m_{\text{Ti}}\sim1\text{g and }\left\vert m_{\text{Be}%
}-m_{\text{Ti}}\right\vert \sim1\mu\text{g.} \label{eq:mass}%
\end{equation}
Assuming that the gravitational and the inertial mass are same, we obtain from
eqn. $\left(  \ref{eq:aaeqn}\right)  $ that%
\begin{equation}
\eta\left(  A,B\right)  =\frac{2\left\vert a_{A}-a_{B}\right\vert }{\left\vert
a_{A}+a_{B}\right\vert }\sim\beta m_{A}\left\vert m_{A}-m_{B}\right\vert
v^{2}\text{.} \label{eqn:eta}%
\end{equation}
Plugging eqn. $\left(  \ref{eq:mass}\right)  $ into eqn. $\left(
\ref{eqn:eta}\right)  $ gives%
\begin{equation}
\beta_{0}^{\text{Be/Ti}}\lesssim1, \label{eq:beta}%
\end{equation}
where we assume $v\sim1$m/s. The superscript Be/Ti in the above equation means
that the upper bound $\beta_{0}$ is for the Be/Ti test bodies. For quarks, it
has been found that\cite{Quesne:2009vc}%
\begin{equation}
\beta_{0}^{q}\approx3^{2}N_{\text{nuc}}^{2}\beta_{0}^{\text{Be/Ti}},
\end{equation}
where $N_{\text{nuc}}$ is the number of nucleons in the test bodies. Since
$N_{\text{nuc}}\sim\frac{10^{-3}\text{kg}}{1.67\times10^{-27}\text{kg}%
}=10^{24}$, it is easy to see that%
\begin{equation}
\beta_{0}^{q}\lesssim10^{49}\text{,}%
\end{equation}
which is much weaker than that from the precession of the perihelion of the Mercury.

Now we shall work out another simple example of a one-dimensional harmonic
oscillator with the potential%
\begin{equation}
V\left(  x\right)  =\frac{m\omega^{2}x^{2}}{2}\text{.}%
\end{equation}
In this example, integrating eqn. $\left(  \ref{eq:Ceqn}\right)  $ gives%
\begin{equation}
\pm\omega\left(  t+C\right)  =\left(  1-\beta mE\right)  \arcsin\left(
\sqrt{\frac{m\omega^{2}}{2E}}x\right)  -\beta mE\sqrt{\frac{m\omega^{2}}{2E}%
}x\sqrt{1-\frac{m\omega^{2}x^{2}}{2E}}, \label{eq:xosc}%
\end{equation}
where the terms higher than $\mathcal{O}\left(  \beta\right)  $ are discarded.
From the LHS of eqn. $\left(  \ref{eq:xosc}\right)  $, one finds that the
oscillator moves between $x=\pm\sqrt{\frac{2E}{m\omega^{2}}}$, which is the
same as in the usual case. Differentiating both sides of eqn. $\left(
\ref{eq:xosc}\right)  $ with respect to $t$ gives the velocity%
\begin{equation}
\frac{dx}{dt}=\frac{\pm\sqrt{\frac{2E}{m}}\sqrt{1-\frac{m\omega^{2}x^{2}}{2E}%
}}{1+2\beta mE\left(  \frac{m\omega^{2}x^{2}}{2E}-1\right)  },
\end{equation}
which implies that $\beta mE<\frac{1}{2}$ otherwise $\frac{dx}{dt}$ would blow
up for some $x\in\left[  -\sqrt{\frac{2E}{m\omega^{2}}},\sqrt{\frac
{2E}{m\omega^{2}}}\right]  $. At $x=\pm\sqrt{\frac{2E}{m\omega^{2}}}$, we have
$\frac{dx}{dt}=0$. Therefore, $x=\pm\sqrt{\frac{2E}{m\omega^{2}}}$ are turning
points and the motion is periodic. Furthermore, eqn. $\left(  \ref{eq:xosc}%
\right)  $ gives that the period of the motion is
\begin{equation}
T=\frac{2\pi}{\omega}\left(  1-\beta mE\right)  .
\end{equation}
Since $\beta>0$, the effects of the minimal length would slow down the
oscillation. Consider a simple gravity pendulum with the mass $m=0.1$kg and
the length $l=1m$, whose period in the usual case is $T_{0}\sim2$s. For the
pendulum, the deformation parameter is%
\begin{equation}
\beta_{0}^{p}\approx\frac{\beta_{0}^{q}}{3^{2}N_{nuc}^{2}}\sim10^{-53}%
\beta_{0}^{q}.
\end{equation}
The correction due to the minimal length to the period is
\[
\left\vert \Delta T\right\vert =T_{0}\beta^{p}mE\sim10^{-55}\beta_{0}%
^{q}\text{s}\lesssim10^{-19}\text{s,}%
\]
where we use $\beta_{0}^{q}<10^{36}$. The correction is too small to observe.

Generally, it can be inferred from the above two examples that the minimal
length correction to some physical quantity $A$ of a non-relativistic
classical system is around%
\begin{equation}
\Delta_{A}=\frac{\left\vert A_{\text{D}}-A_{\text{U}}\right\vert }{\left\vert
A_{\text{U}}\right\vert }\sim\beta m^{2}v^{2},
\end{equation}
where $A_{\text{D}\left(  \text{U}\right)  }$ is the value of $A$ which is
calculated by the deformed theory (usual theory) and $m$ and $v$ are the
typical mass and velocity of the test bodies, respectively. Since $\beta$ is
for the test bodies, we have%
\begin{equation}
\beta\sim\beta_{0}^{q}\left(  \frac{m}{\text{kg}}\right)  ^{-2}\times10^{-54},
\end{equation}
where $\beta_{0}^{q}$ is for quarks. Define $\Delta_{A}^{E}=\frac{\left\vert
A_{\text{U}}-A_{\text{E}}\right\vert }{\left\vert A_{\text{E}}\right\vert }$ ,
where $A_{\text{E}}$ is the value of $A$ measured by the experiment. Thus, we
find%
\begin{equation}
\Delta_{A}\lesssim\Delta_{A}^{E}\Rightarrow\beta_{0}^{q}\lesssim\left(
\frac{v}{\text{m}^{2}/\text{s}^{2}}\right)  ^{-2}\Delta_{A}^{E}\times10^{56}.
\label{eq:betaQ}%
\end{equation}
For the precession of the perihelion of the Mercury, $A$ is the angular
velocity $\omega$ of the Mercury, $\Delta_{\omega}^{E}\sim$ $10^{-11}$ and
$v\sim4\times10^{4}$m$/$s. As a result, we can reproduce the upper bound on
$\beta_{0}^{q}$%
\begin{equation}
\beta_{0}^{q}\lesssim10^{36}.
\end{equation}
It is also interesting to note that the upper bound on $\beta_{0}^{q}$ in eqn.
$\left(  \ref{eq:betaQ}\right)  $ is proportional to $v^{-2}$ and independent
of $m$. In order to put stronger constraints on $\beta_{0}^{q}$, one might
need to look for the experiments with the high typical speed, possible a
relativistic one.

\subsubsection{Statistical Physics in Deformed Spaces}

In statistical physics we often need to calculate sums of the form%
\begin{equation}
\sum\limits_{n}g\left(  \frac{E_{n}}{kT}\right)  , \label{eq:sum}%
\end{equation}
where $E_{n}$ is the energy of $n$-th level, $g$ is some function and $k$ is
the Boltzmann constant. For example, $g\left(  x\right)  =e^{-x}$ for the
partition function of a quantum system obeying the Boltzmann statistics. Now
we consider a $1D$ non-relativistic quantum system under the potential
$V\left(  x\right)  $ in deformed spaces with the deformed commutation
relation $\left(  \ref{eq:generalGUP}\right)  $. For this system, we will use
the Bohr-Sommerfeld quantization condition $\left(  \ref{bsquant}\right)  $ to
show that under the condition $E_{n+1}-E_{n}\ll kT$, the sum in eqn. $\left(
\ref{eq:sum}\right)  \,$\ is equal to%
\begin{equation}
\sum\limits_{n}g\left(  \frac{E_{n}}{kT}\right)  \approx\int\frac{dxdP}%
{2\pi\hbar f\left(  P\right)  }g\left(  \frac{E\left(  P,x\right)  }%
{kT}\right)  , \label{eq:BSinStats}%
\end{equation}
where $E\left(  P,x\right)  =\frac{P^{2}}{2m}+V\left(  x\right)  $.

First, we observe that the integral in eqn. $\left(  \ref{bsquant}\right)  $
is exactly half the area surrounded by the classical trajectory of the
particle in phase space of $p$ and $x$. Thus, we find
\begin{equation}
A_{E_{n}}=\left(  2n+1\right)  \pi\hbar,
\end{equation}
where $E_{n}$ is the energy of $n$-th level and $A_{E_{n}}$ denote the entire
area inside the trajectory corresponding to the energy $E_{n}$. Let us denote
the domain enclosed between the $n$-th and $n+1$-th trajectory as $D_{n}$,
whose area is $A_{E_{n+1}}-A_{E_{n}}=2\pi\hbar$. Thus, we have%
\begin{equation}
\int_{D_{n}}dxdp=2\pi\hbar\text{.}%
\end{equation}
It is easy to see that%
\begin{equation}
g\left(  \frac{E_{n}}{kT}\right)  =g\left(  \frac{E_{n}}{kT}\right)
\int_{D_{n}}\frac{dxdp}{2\pi\hbar}.
\end{equation}
If $E_{n+1}-E_{n}\ll kT$, we find%
\begin{equation}
g\left(  \frac{E_{n}}{kT}\right)  \approx\int_{D_{n}}\frac{dxdp}{2\pi\hbar
}g\left(  \frac{E\left(  P,x\right)  }{kT}\right)  .
\end{equation}
The sum in eqn. $\left(  \ref{eq:sum}\right)  $ becomes%
\begin{equation}
\sum\limits_{n}g\left(  \frac{E_{n}}{kT}\right)  \approx\sum\limits_{n}%
\int_{D_{n}}\frac{dxdp}{2\pi\hbar}g\left(  \frac{E\left(  P,x\right)  }%
{kT}\right)  =\int\frac{dxdp}{2\pi\hbar}g\left(  \frac{E\left(  P,x\right)
}{kT}\right)  ,
\end{equation}
where the integral is over the entire phase space of $x$ and $p$. Using
$\frac{dP\left(  p\right)  }{dp}=f\left(  P\right)  $, we conclude the proof
of eqn. $\left(  \ref{eq:BSinStats}\right)  $. Note that the formula $\left(
\ref{eq:BSinStats}\right)  $ has also been obtained in \cite{Fityo:2008zz} by
calculating the Jacobian $J=\frac{\partial\left(  X,P\right)  }{\partial
\left(  x,p\right)  }$ and in \cite{Wang:2010ct,Chang:2001bm} by using
Liouville theorem.

Now consider a $1D$ harmonic oscillator with the potential $V\left(  x\right)
=\frac{m\omega^{2}x^{2}}{2}$. For $T\gg\frac{E_{n+1}-E_{n}}{k}\sim\frac
{\hbar\omega}{k}$, the partition function for the oscillator is
\begin{align}
Z  &  =\int\frac{dxdP}{2\pi\hbar f\left(  P\right)  }\exp\left[  -\frac
{1}{2kT}\left(  \frac{P^{2}}{m}+m\omega^{2}x^{2}\right)  \right] \nonumber\\
&  =\frac{kT}{\hbar\omega\sqrt{\pi}}\int_{-\infty}^{\infty}\frac{dx}{f\left(
\sqrt{2kTm}x\right)  }\exp\left(  -x^{2}\right)  ,
\end{align}
where $x=\frac{P}{\sqrt{2kTm}}$. Suppose $f\left(  P\right)  =1+\beta P^{2}$.
The partition function becomes%
\begin{equation}
Z=\frac{kT}{\hbar\omega\sqrt{\pi}}\int_{-\infty}^{\infty}\frac{dx}{1+2kTm\beta
x^{2}}\exp\left(  -x^{2}\right)  \text{.}%
\end{equation}
If $kTm\beta\ll1$, we find
\begin{equation}
Z\approx\frac{kT}{\hbar\omega}\left(  1-kTm\beta\right)  \text{.}%
\end{equation}
If $kTm\beta\gg1$, we find%
\begin{equation}
Z=\frac{kT}{\hbar\omega\sqrt{\pi}}\frac{1}{\sqrt{2kTm\beta}}\int_{-\infty
}^{\infty}\frac{dy}{1+y^{2}}\exp\left(  -\frac{y^{2}}{2kTm\beta}\right)
\approx\frac{\sqrt{kT\pi}}{\sqrt{2m\beta}\hbar\omega},
\end{equation}
where $y=\sqrt{2kTm\beta}x$. The average energy of the oscillator is%
\begin{equation}
\bar{E}=-\frac{\partial}{\partial\left(  \frac{1}{kT}\right)  }\ln
Z\approx\left\{
\begin{array}
[c]{c}%
kT\left(  1-\beta mkT\right)  \text{ \ for }kTm\beta\ll1\text{\ }\\
\text{ \ \ }\frac{1}{2}kT\,\text{\ \ \ \ \ \ \ for }kTm\beta\gg1
\end{array}
\right.  . \label{eq:averageE}%
\end{equation}
If $kTm\beta\gg1$, the energy for one degrees of freedom in the equipartition
theorem in deformed spaces is only half of that in the usual case.

Einstein assumed that the atoms in a crystal is equivalent to $3N$ harmonic
oscillators and calculated heat capacities of solids. For an atom with the
standard atomic weight $A_{r}$, we find%
\begin{equation}
\frac{1}{km\beta^{a}}\sim10^{18}\left(  \frac{A_{r}}{100}\right)  \left(
\frac{10^{36}}{\beta_{0}^{q}}\right)  K\text{.}%
\end{equation}
Usually heat capacities of solids are measured at $T\sim10^{2}K$. In this
case, eqn. $\left(  \ref{eq:averageE}\right)  $ gives that the molar specific
heat of a solid for $\frac{\hbar\omega}{k}\ll T\ll\frac{1}{km\beta^{a}}$ is%
\begin{equation}
C=3R\left(  1-\beta^{a}mkT\right)  ,
\end{equation}
where $R=8.31$JK$^{-1}$mol$^{-1}$ is the gas constant. For a solid consisting
of atoms with $A_{r}$, we have%
\begin{equation}
\Delta C=\left\vert \frac{C-3R}{3R}\right\vert \sim10^{-52}\beta_{0}%
^{q}\left(  \frac{100}{A_{r}}\right)  \left(  \frac{T}{100K}\right)  .
\end{equation}
For example, the heat capacity of Tungsten at $T=400K$ is $C=24.92$JK$^{-1}%
$mol$^{-1}$\cite{CRC}, which implies%
\begin{equation}
\beta_{0}^{q}\lesssim10^{50}\text{.}%
\end{equation}

\section{Conclusions}

\label{sec:conclusion}

In this paper, we considered a homogeneous field in the deformed quantum
mechanics with minimal length. The physical motivation for this is to obtain
the WKB connection formula and prove the Bohr-Sommerfeld quantization rule
rigorously in the deformed quantum mechanics. By studying the leading
asymptotic behavior of the physically acceptable wave function in the physical
region, we found the contour for its integral representation. Through the
integral representation, the asymptotic expansions of the physically
acceptable wave function at both large positive and large negative values of
$\rho$ were given.

We then used the obtained asymptotic expansions to get the WKB connection
formula, which proceeds from classically forbidden region to classically
allowed one through a smooth turning point, and had the Bohr-Sommerfeld
quantization rule proved rigorously up to $\mathcal{O}\left(  \beta
^{2}\right)  $. A new interesting feature appearing in the presence of
deformation was that our WKB connection formula does not work for a sharp
turning point. The connection through such a point might need a higher order
approximation to the potential near it.

We discussed the competition between the quantum gravity correction on the
first order WKB approximation and the second order WKB approximation. If the
former is not overwhelmed by the latter, the energy of the particle considered
should be high enough according to (\ref{betavshconstraint}). We also showed
that, if the energy levels $E_{n}^{\left(  0\right)  }$ of a bound state are
given in the ordinary quantum mechanics, the deformed energy levels are%
\begin{equation}
E_{n}=E_{n}^{\left(  0\right)  }+\beta E_{n,0}^{1}+\mathcal{O}\left(
\hbar\right)  \mathcal{O}\left(  \beta\right)  +\mathcal{O}\left(  \beta
^{2}\right)  ,
\end{equation}
where $\beta E_{n,0}^{1}$ is the $\mathcal{O}\left(  \beta\right)  $ quantum
gravity correction on the first order WKB approximation. Finally, we used the
Hamilton-Jacobi method to study the effects of the minimal length on the
classical motions. The Bohr-Sommerfeld quantization was then used to
investigate statistical physics in deformed spaces with the minimal length.
Upper bounds on $\beta_{0}^{q}$ were obtained from measurements of the
\textquotedblleft Eotvos ratio\textquotedblright\ and the heat capacity of Tungsten.

\bigskip\bigskip

\textbf{Conflict of Interest}

The authors declare that there is no conflict of interests regarding the
publication of this paper.

\begin{acknowledgments}
We are grateful to Benrong Mu, Houwen Wu and Zheng Sun for useful discussions.
This work is supported in part by NSFC (Grant No. 11175039, 11147106, 11005016
and 11375121) and the Fundamental Research Funds for the Central Universities.
\end{acknowledgments}

\end{document}